\documentclass[11pt,twoside]{article}

\usepackage{asp2006}
\usepackage{graphicx}
\usepackage{lscape}
\usepackage{natbib}
\usepackage{amsmath}

\usepackage{wasysym} 

\begin{document}

\title{The very high energy $\gamma$-ray view of the Galactic Centre as of early 2010}

\author{Christopher van Eldik for the H.E.S.S. Collaboration}
\affil{Max-Planck-Institut f\"ur Kernphysik, P.O. Box 103980, D-69029 Heidelberg, Germany}

\begin{abstract}
Progress in the Imaging Atmospheric Cherenkov Technique has enabled first
sensitive observations of the innermost few 100~pc of the Milky Way in
Very High Energy (VHE; $>100$~GeV) $\gamma$-rays. Observations by the H.E.S.S. instrument deliver the at date most precise data on this peculiar region, and provide an interesting view onto the acceleration and propagation of energetic particles near the Galactic Centre. Besides two point-like sources -- one coincident with the supermassive black hole (SMBH) Sgr~A* -- diffuse VHE emission has been discovered within a $1^\circ$ region around the center.  The current VHE $\gamma$-ray view of the region is reviewed, and possible counterparts of the $\gamma$-ray sources and the origin of the diffuse emission are discussed.
\end{abstract}

\section{Introduction}
During the last decades, the quality of astronomical data from the Galactic Center (GC) region has increased dramatically. Nowadays, the GC is routinely monitored at radio, infrared, X-ray, and hard X-ray/soft $\gamma$-ray energies, and the distribution of dust and atomic and molecular material is surveyed with increasing accuracy. Altogether, these efforts provide a rich multi-wavelength data set of this unique and complex region of the sky, which -- due to its proximity to Earth -- is an ideal laboratory for investigating the astrophysics of galactic nuclei in general.

The radio picture \citep{LaRosa00} of the inner few 100~pc around the gravitational centre reveals numerous sources of non-thermal radiation, most probably caused by synchrotron radiation of relativistic electrons. Particle acceleration to supra-thermal energies is therefore believed to take place at the GC, possibly to a few 10~TeV and beyond. Therefore, this region is a prime target for observations at very high energies, since particles of TeV energy, upon interacting with background photons or molecular material, produce $\gamma$-rays in the VHE range. These $\gamma$-rays, which pass the GC dust torus and Galactic magnetic fields unaffected, are an excellent tracer for sites of particle acceleration to highest energies.

The VHE $\gamma$-ray flux arriving at Earth from typical sources is low. From the direction of the GC point source, a fairly strong $\gamma$-ray emitter, a flux well below 1~m$^{-2}$~yr$^{-1}$ is recorded for $\gamma$-rays with energies above 1~TeV. However, the a large detection area available to ground-based VHE $\gamma$-ray instruments guarantees good $\gamma$-ray statistics, given typical observation times of 1-50 hours. Although VHE $\gamma$-rays are efficiently absorbed in the atmosphere, they can be detected on ground by means of the Cherenkov light of the relativistic air shower particles that are produced in the absorption process \citep[see, e.g.,][for a recent review]{2008RPPh...71i6901A}. The light pool covers an area of about 50.000~m$^2$ on the ground, and instruments like H.E.S.S., MAGIC, VERITAS, and Cangaroo-III use (arrays of) large mirror telescopes to image the Cherenkov light onto sensitive segmented cameras. Detection of VHE $\gamma$-rays suffers from background caused by the much more numerous charged cosmic rays impinging Earth's atmosphere. This background can, however, be efficiently suppressed by stereoscopic observation and analysis of the shape of the recorded images. During the last years, Imaging Atmospheric Cherenkov Telescopes (IACTs) opened up a new observational window to the universe: more than 80 VHE $\gamma$-ray sources -- both Galactic and extragalactic -- have been discovered since then, and at least 6 source classes identified \citep{Hinton:2009}.

The H.E.S.S. instrument provides the at date best sensitivity for GC observations. The telescope array consists of four 13~m diameter IACTs, located in the Khomas Highlands of Namibia, roughly $23^\circ$ south of the equator, where the GC culminates close to zenith during the summer months, providing ideal observation conditions. With its $\sim 5^\circ$ field-of-view H.E.S.S. is able to observe a $\sim 600$~pc region around the GC with a single pointing of the instrument (assuming a distance to the GC of 8~kpc). Besides strong VHE point-source emission from the direction of Sgr~A* (discussed in sections \ref{sec:Discovery} and \ref{sec:HGC}), H.E.S.S. detected $\gamma$-rays from a pulsar wind nebula (PWN) inside the shell of G~0.9+0.1, a well-known supernova remnant. A comprehensive discussion about G~0.9+0.1 is beyond the scope of this report. Details can be found in \citet{Aharonian:2005br}. Besides the two point sources, H.E.S.S. discovered diffuse VHE emission along the Galactic Centre ridge, correlated with the distribution of molecular clouds in a region of diameter $\sim 300$~pc around the GC. This diffuse emission and its implications are discussed in section \ref{sec:Diffuse}

\section{Discovery of a strong VHE point source at the GC}
\label{sec:Discovery}

Given the importance of the GC as a possible multi-TeV particle accelerator, the region was in the focus of IACTs since the early days of this detection technique. It took, however, until 2004 that a VHE $\gamma$-ray signal was detected from the GC by three instruments almost simultaneously \citep{Tsuchiya:2004wv,Kosack:2004ri,Aharonian:2004wa}. First results were, however, at odds with one another: while the experiments agreed that the emission was point-like, and no significant flux variability was detected from the source, there was disagreement on the power-law spectral indexes and flux normalisations of the energy spectra measured. The Cangaroo-II instrument reported a $10~\sigma$ detection above 250~GeV and a very steep spectrum, with an index of $4.6\pm 0.5$ \citep{Tsuchiya:2004wv} and a flux normalisation at
1~TeV of about $2.7\times 10^{-12}$~cm$^{-2}$~s$^{-1}$~TeV$^{-1}$. The Whipple instrument detected the GC
with a marginal significance of 3.7~$\sigma$ above the background. The
integral $\gamma$-ray flux reported was $(1.6\pm 0.5_{\mathrm{stat}} \pm
0.3_{\mathrm{sys}})\times 10^{-12}$~cm$^{-2}$~s$^{-1}$ above 2.8~TeV,
roughly two orders of magnitude larger than the flux measured by
Cangaroo-II at these energies. 

17 hours of observations during 2003 with two telescopes of the partially completed H.E.S.S. array resulted in two independent data sets and two clear detections of the GC source (with significances above the background of $6.1~\sigma$ and $9.2~\sigma$), henceforth called HESS~J1745-290 \citep{Aharonian:2004wa}. The energy spectra of these two measurements were compatible, and hard spectral indexes were reported. However, the results were significantly different from those of Whipple and Cangaroo-II both in terms of spectral index and flux normalisation.
49~hours of follow-up observations carried out in 2004 with the completed H.E.S.S. array 
confirmed, however, the early H.E.S.S. results: from a power-law fit of the 2004 data, a photon index of $\Gamma=2.25\pm 0.04_\mathrm{stat}\pm 0.10_\mathrm{sys}$ and an integral flux
above 1~TeV of $(1.87\pm 0.10_\mathrm{stat} \pm
0.30_\mathrm{sys}) \cdot 10^{-12}$~cm$^{-2}$~s$^{-1}$ was obtained
\citep{Aharonian:2006wh}. As for the earlier measurements, the the $\gamma$-ray emission was found point-like and coincident within the errors with the position of Sgr~A*. In 2004 and 2005 the MAGIC collaboration observed HESS~J1745-290 and also reported the detection of a point-like, non-variable source \citep{Albert:2005kh}. These measurements confirmed the hard spectrum found by H.E.S.S., with consistent flux levels. 

At these days, the differences between the various measurements could either be explained by
rapidly varying $\gamma$-ray emission from the source (with the caveat that none of the
experiments detected significant variability in its own data set),
or some hidden systematics in the data analyses. Indeed, after a careful
reanalysis of the Whipple data \citep{Kosack2005} the flux level was
corrected, and a differential energy spectrum
matching the H.E.S.S. and MAGIC spectra was obtained. Moreover,
observations with the CANGAROO-III array recently yielded a
differential energy spectrum consistent with the H.E.S.S. and MAGIC
results \citep{Mizukami2008}, such that the initial disagreements about the spectral properties of HESS~J1745-290 seem to be settled.

\begin{figure}[!ht]
\includegraphics[width=0.75\textwidth]{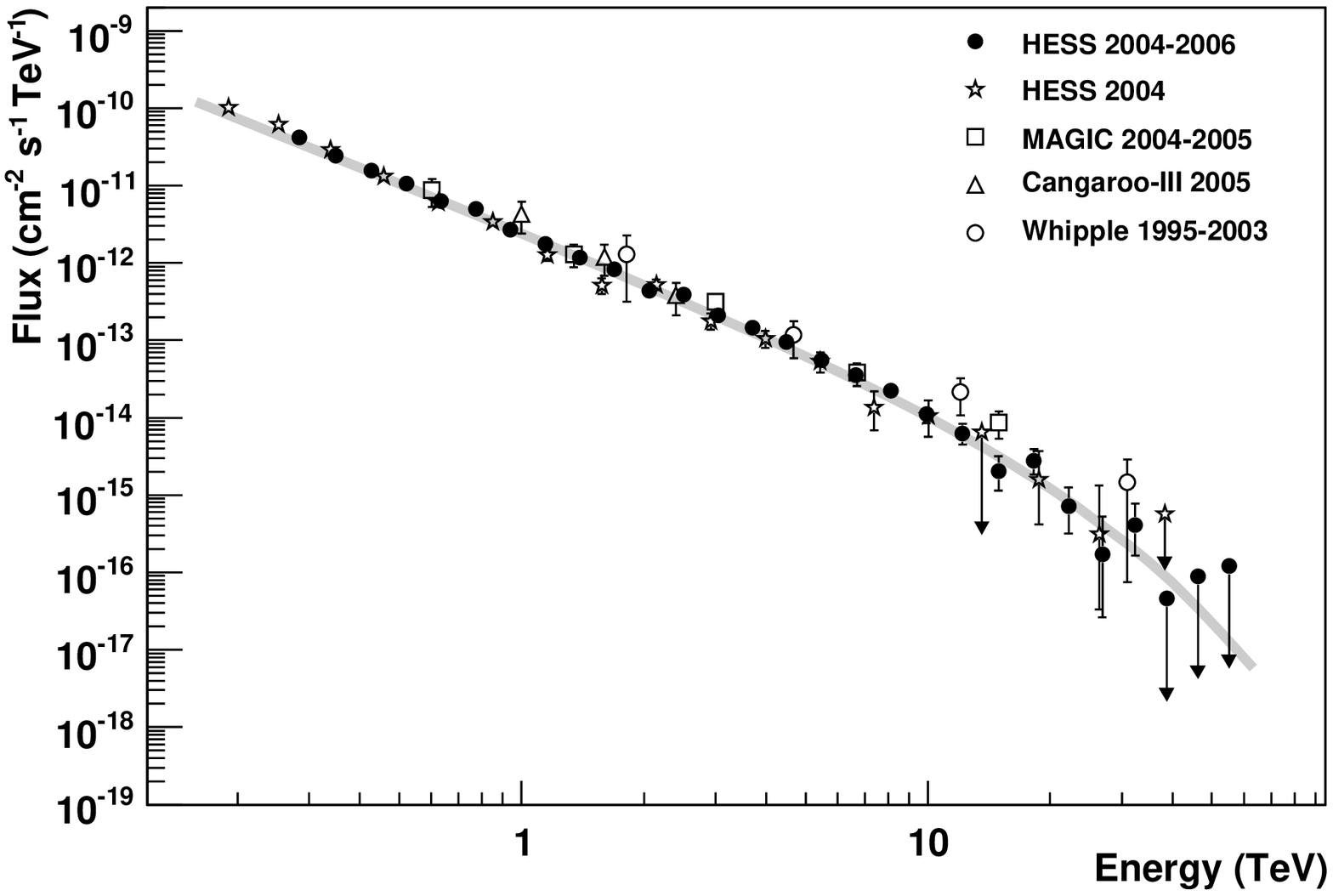}
\label{fig:Spectra}
\caption{Compilation of VHE $\gamma$-ray spectra
of the GC source HESS~J1745-290. Data points are
taken from \citet{Kosack2005}, \citet{Aharonian:2006wh}, \citet{Albert:2005kh}, \citet{Mizukami2008}, and \citet{GCSpectrum}. 
The curve shows a power-law fit with exponential cut-off to the most recent H.E.S.S. data \citep[see text and][]{GCSpectrum}. Note that the H.E.S.S. spectra were corrected for a flux contribution of $\sim 15$\% from diffuse emission. Upper limits are given at 95\% CL. The early results from Whipple \citep{Kosack:2004ri} and Cangaroo-II \citep{Tsuchiya:2004wv} are not shown.}
\end{figure}

An update on the spectrum of the source was recently put forward by the H.E.S.S. collaboration. Based on 93 hours (live time) of observations during the years 2004-2006, a clear deviation of the energy spectrum from a single power-law distribution is observed for the first time. The spectrum is well described by a power-law with exponential cut-off,
\begin{equation*}
\frac{dN}{dE} = \Phi_0 \cdot \left( \frac{E}{1\mathrm{TeV}} \right)^{-\Gamma} e^{-\frac{E}{E_\mathrm{c}}},
\end{equation*}
with $\Phi_0 = (2.40\pm 0.10)\cdot 10^{-12}$~TeV$^{-1}$cm$^{-2}$s$^{-1}$, $\Gamma=2.10\pm 0.04$, and $E_\mathrm{c}=(14.70\pm 3.41)$~TeV \citep{GCSpectrum}. A single power-law fit is rejected with a $\chi^2/\mathrm{dof}$ of $64/27$, but the data are equally well described by a smoothed broken power-law \citep{GCSpectrum}. Fig. \ref{fig:Spectra} shows a compilation of the at date available VHE $\gamma$-ray flux measurements of HESS~J1745-290, together with the above given fit to the latest H.E.S.S. data,
indicating the recent agreement between the different instruments. 

\section{HESS~J1745-290: a prime example of an unidentified $\gamma$-ray source}
\label{sec:HGC}
Despite the recent progress in obtaining a consistent picture of the GC VHE emission, the actual mechanism that produces the emission is not yet understood. A firm identification is particularly hampered by the --
compared to radio or X-ray instruments -- modest angular resolution of
the current generation of Cherenkov telescopes ($\leq 5'$ for a single
$\gamma$-ray at TeV energies). Compared to the projected distances of counterpart candidates in the GC, the VHE emission region is relatively large, giving rise to source confusion in this densely populated part of the Galaxy. 
Nevertheless can VHE $\gamma$-ray observations put constraints on counterparts and emission models in various
ways. Without being in conflict with measurements at longer
wavelengths, models for HESS~J1745-290 must explain the following source properties:

\begin{itemize}
\item The energy spectrum between 160~GeV and 70~TeV can be
  characterised by a power-law with exponential cut-off, or a smoothed broken power-law (see section \ref{sec:Discovery}). The integral flux above 1~TeV is $2\cdot 10^{-12}$~cm$^{-2}$~s$^{-1}$. This implies a $\gamma$-ray luminosity of about $10^{35}$~erg~s$^{-1}$ in the 1-10~TeV range. 
\item There is no hint for significant flux variability on any timescale from minutes to years \citep{GCSpectrum}.
\item The centroid of HESS~J1745-290 is, within $8''\pm 9''_\mathrm{stat} \pm 9''_\mathrm{sys}$, coincident with the position of Sgr~A* \citep{GCPositionPaper}, and the intrinsic size of the source amounts to less than 1.2 arc minutes.
\end{itemize}

Although an association of HESS~J1745-290 with Sgr~A* is compelling (and certainly viable in terms of energetics, position and spectrum, see below), there are at least two other objects in direct vicinity of the SMBH which are good candidates for producing the observed $\gamma$-ray flux in parts or in total: the SNR Sgr~A~East and the recently discovered PWN candidate G359.95-0.04.

\subsubsection{The case of Sgr~A~East}
The existence of synchrotron radiation, i.e. the presence of relativistic electrons,
and a large magnetic field \citep[$\approx 2-4$~mG,][]{Yusef96} make Sgr~A~East a compelling
candidate for $\gamma$-ray emission at VHE energies. The energy spectra at radio, X-ray, and soft $\gamma$-ray energies match a modelq in which protons get accelerated in shock waves to an energy of at least 100~GeV \citep{Fatuzzo2003}. Adopting a 4~mG magnetic field, \citet{Crocker2005} estimate a maximum proton energy of $10^{19}$~eV achievable in the Sgr~A~East blast wave. Furthermore, the observed absence of flux variability from HESS~J1745-290 is naturally supported by the fact that particle acceleration is supposed to take place in the extended shell of the SNR.

\begin{figure}[!ht]
\includegraphics[width=0.65\textwidth]{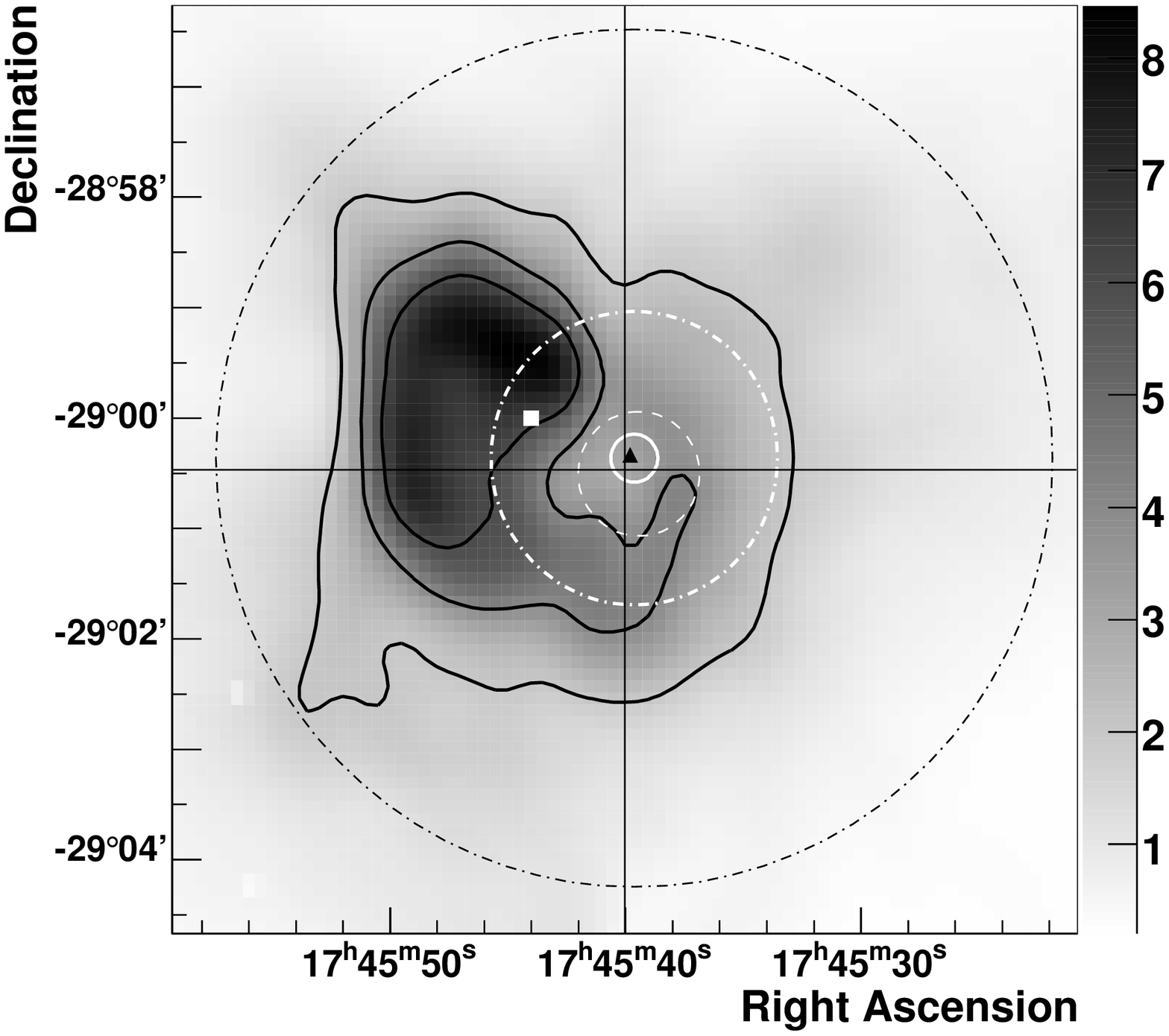}
\caption{90~cm VLA radio flux density map \citep{LaRosa00} of the
    innermost 20~pc of the GC, showing emission Sgr~A~East.
    Black contours denote radio flux levels of 2, 4, and 6~Jy/beam.
    The centre of the SNR \citep{Green:2009qf} is marked by the white
    square, and the positions of Sgr~A~Star \citep{saga_radio} and G395.95-0.04
    \citep[head position,][]{Wang:2005ya} are given by the cross hairs
    and the black triangle, respectively. The 68\% CL total error contour of
    the best-fit centroid position of HESS~J1745-290 is given by the white
    circle. The dashed white circle shows the same contour for the
    previously reported H.E.S.S. measurement \citep{Aharonian:2006wh}.
    The white and black dashed-dotted lines show the 95\% CL upper limit
    contour of the source extension and the 68\% containment region of
    the H.E.S.S. PSF, respectively.}
\label{fig:RadioMap}
\end{figure}

Sgr~A~East was, however, recently excluded as the main counterpart of the VHE emission by arguments of the position of the VHE emission centroid \citep{GCPositionPaper}. For illustration, fig.~\ref{fig:RadioMap} shows a VLA 90~cm image of the innermost 20~pc region of the GC, centred on Sgr~A*. The shell-like radio structure of  Sgr~A~East is
clearly visible. It surrounds Sgr~A* in projection, and its radio emission maximum is only 1.5' (or about 3.5~pc)
away from the position of Sgr~A*, well within the point spread function of the H.E.S.S. measurement. The centroid of HESS~J1745-290, however, with a 68\% CL total error radius of 13'' only, is located in a region where the radio emission from Sgr~A~East is comparatively low. This position measurement is the most precise so far in VHE $\gamma$-ray astronomy, and was achieved after a careful investigation of the pointing systematics of the H.E.S.S. telescopes, reducing the systematic error on the centroid position from 20'' \citep{Aharonian:2006wh} to 6'' per axis. Due to such small errors, Sgr~A~East is ruled out as the bulk emitter of the VHE $\gamma$-rays \citep[at a significance of $3.9~\sigma$ for the most conservative analysis,][]{GCPositionPaper}).

\subsubsection{HESS~J1745-290: a pulsar wind nebula?}
The recent detection of the PWN candidate G359.95-0.04 in a deep Chandra
exposure of the GC region \citep{Wang:2005ya} very much complicates
the identification of HESS~J1745-290. G359.95-0.04 is located only $8.7''$ in
projection (or 0.3~pc) away from Sgr~A*, rendering a discrimination of the two by
means of a position measurement of HESS~J1745-290 impossible (see. Fig. \ref{fig:RadioMap}).  
G359.95-0.04 is rather faint at X-ray energies, with an implied luminosity
of $10^{34}$~erg~s$^{-1}$ in the 2-10~keV band
\citep{Wang:2005ya}, yet about four times brighter than Sgr~A*. It
shows a cometary shape and exhibits a hard 
and non-thermal spectrum which gradually softens when going away from
the ``head'' of the PWN, where the yet undiscovered pulsar is believed
to be located. No radio counterpart of the PWN is found.

Numerical calculations show that a
population of non-thermal electrons can naturally explain both the
X-ray emission of G359.95-0.04 and the VHE $\gamma$-ray emission 
of HESS~J1745-290 \citep{Hinton:2006zk}. Compared to other locations in the
galactic disk the GC region is special because of its dense
radiation fields. TeV electrons up-scatter predominantly the far-IR component
of the radiation field to TeV energies. This
provides roughly an order of magnitude larger luminosity in the
1-10~TeV $\gamma$-ray band than in the 2-10~keV X-ray domain.

\subsubsection{Gamma-ray emission scenarios involving Sgr~A*}
Its low bolometric luminosity ($< 10^{-8} L_{\mathrm{Edd}}$ in the
range from millimetre to optical wavelengths) renders Sgr~A* an
unusually quiet representative of galactic nuclei. At the same time,
this property makes the immediate vicinity of the SMBH transparent to
VHE $\gamma$-rays. \citet{Aharonian:2005ti} show that the
absence of dense IR radiation fields enables photons with an
energy of up to several TeV to escape almost unabsorbed from regions
as close as several Schwarzschild radii from the centre of the SMBH.
Therefore, VHE $\gamma$-ray emission produced close to the event
horizon of Sgr~A* provides a unique opportunity to study particle
acceleration and radiation in the vicinity of a black hole. 

Sgr~A* offers several possibilities to produce the observed VHE $\gamma$-ray flux, 
depending on the type of particles accelerated, the model of
acceleration, and finally the  
interaction of the accelerated particles with the ambient magnetic
field or matter. Common scenarios, which do not contradict the
emission at longer wavelengths, include $\gamma$-ray production
close to the SMBH itself \citep{Aharonian:2005ti}, within an ${\cal O}(10)$~pc zone
around Sgr~A* due to the interaction of run-away protons with the
ambient medium \citep{Aharonian:2005b,Liu2006a,Wang:2009}, or electron acceleration 
in termination shocks driven by winds emerging from within a couple of Schwarzschild
radii \citep{Atoyan2004}.

While some of these models suggest correlated
multi-wavelength variability, others predict a steady VHE flux because of diffusion of accelerated particles into the surroundings or acceleration in an extended region far away from the SMBH surface. Therefore, non-observation of variability does not rule out Sgr~A* as a counterpart candidate. On the other hand, the detection of variability in the
VHE data would immediately point to $\gamma$-ray production in its vicinity. The most convincing signature would be the discovery of
correlated flaring in X-rays (or NIR) and VHE $\gamma$-rays. Such a search has been carried out \citep{2008A&A...492L..25A}. In a coordinated multi-wavelength campaign both Chandra and H.E.S.S. observed the GC region, when a major (factor 9 increase) X-ray outburst was detected. During this 1600~s flare the VHE $\gamma$-ray flux stayed constant within errors, and a 99\% CL upper limit on a doubling 
of the VHE flux was derived. When interpreting the X-ray flare as synchrotron emission of TeV electrons, the non-observation of an inverse Compton VHE counterpart implies a magnetic field strength in the emission region of $>50$~mG \citep{2008A&A...492L..25A}. So far no strong constraint can be derived from these findings, because even larger magnetic fields are expected close to Sgr~A*.

\subsubsection{Dark Matter annihilation close to the GC?}
Besides being of astrophysical origin, the observed TeV flux could
potentially stem from annihilation of dark-matter (DM) particles,
which are believed to cluster in a compact cusp around Sgr A*
\citep[e.g.][]{Bergstroem2000}. Halo density profiles are believed to scale
with the radius $r$ like $r^{-\alpha}$, with $\alpha$ between 1
\citep{Navarro1997} and 1.5 \citep{Moore1999} in the most common
models. The fact that HESS~J1745-290 is point-like -- after having accounted for
the underlying diffuse emission -- translates into $\alpha>1.2$,
i.e. a cuspy halo is favoured by the observations \citep{Aharonian:2006wh}. 

Predicted energy spectra for $\gamma$-rays produced in cascade decays
of DM particles, such as MSSM neutralinos or Kaluza-Klein particles, can be
compared to the VHE observations. These spectra are usually curved
both at high energies -- for reasons of energy conservation --, and low
energies, somewhat in disagreement with the observations
(Fig. \ref{fig:Spectra}, see also
\citealt{Aharonian:2006wh}). Furthermore, unusually large DM particle  
masses have to be assumed to account for the fact that the $\gamma$-ray
spectrum extends far beyond 10~TeV. 

The $\gamma$-ray emission from HESS~J1745-290 is therefore not compatible with 
being dominantly produced in the framework of the most common DM
scenarios. As a consequence, the bulk of the 
$\gamma$-ray excess is probably of astrophysics rather than of particle
physics origin. However, an ${\cal O}(10\%)$ admixture of $\gamma$-rays from DM 
annihilations in the signal cannot be ruled out. 
Assuming an NFW-type \citep{Navarro1997} halo profile, 99\% CL upper
limits on the velocity-weighted annihilation cross section $<\sigma
v>$ are at least two orders of magnitude above theoretical
expectations, and thus are not able to put
constraints on current DM model predictions \citep{Aharonian:2006wh}. 

\section{Diffuse \boldmath{$\gamma$}-ray emission: a cosmic ray accelerator at the GC?}
\label{sec:Diffuse}
The diffuse emission detected by H.E.S.S. covers a region of roughly  
$2^\circ$ in galactic longitude ($l$) with an rms width of about $0.2^\circ$
in galactic latitude ($b$). The reconstructed $\gamma$-ray spectrum
integrated within $|l|\leq 0.8$ and $|b|\leq 0.3$ is well-described by
a power law with photon index $\Gamma=2.29$ \citep{Aharonian:2006au},
similar to what is observed for HESS~J1745-290. There is, at least for $|l|\leq 1^\circ$, a
strong correlation between the morphology of the observed
$\gamma$-rays and the density of molecular clouds \citep[traced by CS 
emission,][]{cs}. This is a strong
indication for the presence of an accelerator of (hadronic) cosmic
rays in the GC region, since the energetic hadrons would
interact with the material in the clouds, giving rise to the observed
$\gamma$-ray flux via $\pi^0\to\gamma\gamma$ decays. The idea of local acceleration
is further supported by the fact that the
measured $\gamma$-ray flux is both larger and harder than expected in a
scenario where the molecular material is only bathened in a sea of Galactic
cosmic rays of similar properties as measured in our solar
neighbourhood. 
A distribution of electron accelerators, such as PWNe, that cluster
similarly to the gas distribution, has also been discussed
\citep[e.g.][]{Aharonian:2006au,Wommer2008}. Given the ${\cal O}(100\mathrm{\mu}$G)
magnetic fields in the region, electrons of several TeV energy would,
however, rapidly cool via synchrotron radiation, such that instead of diffuse emission several point-like VHE
$\gamma$-ray sources would be expected. 

In the context of identifying the accelerator, the fact
that no emission is seen farther away than $|l|\approx 1^\circ$ might be
particularly important. A simple, yet convincing explanation is that the cosmic rays were
accelerated in a rather young source near the very centre of
the galaxy, and underwent diffusion away from the
accelerator into the surrounding medium. Assuming a typical diffusion
coefficient of $10^{30}$~cm$^2$~s$^{-1}$ (or 3~kpc$^2$~Myr$^{-1}$) for
TeV protons in the Galactic disk, a source age of about $10^4$ years
can reproduce the observed $\gamma$-ray morphology
\citep{Aharonian:2006au}, in particular the lack of emission beyond
$1^\circ$ distance from the centre.  

\citet{Buesching2007} follow up on this idea.
Starting from a source of non-thermal protons at the GC
and the fairly well-known distribution of molecular material,
the authors model the $\gamma$-ray flux from the region in a time dependent
diffusion picture. Neglecting a possible energy dependence of the
diffusion process, they compute the diffusion coefficient for which the H.E.S.S.
results are matched best, for a variety of source ages and source on-times. 
In a similar approach, using a more accurate 3d model for the distribution of molecular clouds, \citet{2009APh....31...13D} obtain a best-fit diffusion coefficient
of 3~kpc$^2$~Myr$^{-1}$, close to what was initially suggested \citep{Aharonian:2006au}. Scaling the
diffusion coefficient $k$ with the cosmic ray rigidity $\zeta$, $k=k_0
(\zeta/\zeta_0)^{0.6}$, $\zeta_0 = 1$~GV$/c$, B\"usching et al. find a
value of $k_0$ which is significantly smaller than the local value,
suggesting enhanced turbulence and larger magnetic fields than in the
solar neighbourhood. 

\citet{Buesching2008} try to explain
both the diffuse emission and the point-source HESS~J1745-290 within a single
model. The authors assume that the cosmic rays responsible for the diffuse emission were
accelerated in the shock wave of Sgr~A~East 5-10~kyr ago, but
acceleration stopped well before the present time. At some point, the
shock wave of Sgr~A~East collided with Sgr~A*, such that particle acceleration near
the SMBH was initiated, leading to the observed VHE $\gamma$-ray emission from
HESS~J1745-290. Assuming that the diffusion coefficient found for the diffuse
emission is also valid close to the SMBH, this last round of particle
acceleration can only have happened in the recent past
(${\cal O}(100)$~yr) to be consistent with the point-like morphology of HESS~J1745-290. 

It should, however, be noted that there are other processes which can
explain the emission from HESS~J1745-290 (see section \ref{sec:HGC}). Furthermore, recent
simulations may indicate that the diffuse emission might be better
explained by inter-cloud acceleration of cosmic rays via the Fermi-II
process \citep{Wommer2008}. More sensitive observations are needed to
ultimately prove which of the discussed scenarios of the VHE $\gamma$-ray view
of the GC is correct.

\section{Conclusions}
Six years after the discovery of VHE $\gamma$-ray emission from the
direction of the GC, observations with Imaging Atmospheric Cherenkov
Telescopes provide a very sensitive view of this interesting region.
With the recent data from the H.E.S.S. instrument, a rich VHE $\gamma$-ray
morphology becomes evident, giving strong indication for the existence of
a cosmic ray accelerator within the central 10~pc of the Milky Way.

A strong $\gamma$-ray point source is found coincident with the position of Sgr~A*, 
within unprecedentedly small errors. Source confusion near the GC make a solid
identification still difficult, given the non-observation of variability and the moderate angular
resolution of current IACTs. However, the recent progress in
improving on the systematic errors of the centroid position of
HESS~J1745-290 excludes the SNR Sgr~A~East as the dominant
source of the $\gamma$-ray emission. A major contribution from the
annihilation of DM particles is also excluded, based on the shape of the $\gamma$-ray energy spectra.

Future observations with even more
sensitive instruments such as CTA will significantly advance
our knowledge about the GC region at VHE energies. The
recently launched Fermi satellite will extend the energy range
down to less than 100~MeV, such that unbroken sensitivity coverage will
be provided over 6 orders of magnitude in energy.

\acknowledgements The author would like to thank the organisers for
the opportunity to present this overview at the workshop. The support of the Namibian authorities and of the University of
  Namibia in facilitating the construction and operation of H.E.S.S.
  is gratefully acknowledged, as is the support by the German Ministry
  for Education and Research (BMBF), the Max Planck Society, the
  French Ministry for Research, the CNRS-IN2P3 and the Astroparticle
  Interdisciplinary Programme of the CNRS, the U.K. Science and
  Technology Facilities Council (STFC), the IPNP of the Charles
  University, the Polish Ministry of Science and Higher Education, the
  South African Department of Science and Technology and National
  Research Foundation, and by the University of Namibia. We appreciate
  the excellent work of the technical support staff in Berlin, Durham,
  Hamburg, Heidelberg, Palaiseau, Paris, Saclay, and in Namibia in the
  construction and operation of the equipment.  

\bibliographystyle{mn2e}
\bibliography{vaneldik_GC_review}

\begin{thebibliography}{}

\bibitem[\protect\citeauthoryear{{Acero} et~al.,}{{Acero}
  et~al.}{2009}]{GCPositionPaper}
{Acero} F.,  et~al., 2009, Mon. Not. R. Astron. Soc., ?, ?

\bibitem[\protect\citeauthoryear{{Aharonian}, {Buckley}, {Kifune} \&
  {Sinnis}}{{Aharonian} et~al.}{2008}]{2008RPPh...71i6901A}
{Aharonian} F.,  {Buckley} J.,  {Kifune} T.,    {Sinnis} G.,  2008, Reports on
  Progress in Physics, 71, 096901

\bibitem[\protect\citeauthoryear{Aharonian et~al.,}{Aharonian
  et~al.}{2004}]{Aharonian:2004wa}
Aharonian F.,  et~al., 2004, Astron. Astrophys., 425, L13

\bibitem[\protect\citeauthoryear{Aharonian et~al.,}{Aharonian
  et~al.}{2005}]{Aharonian:2005br}
Aharonian F.,  et~al., 2005, Astron. Astrophys., 432, L25

\bibitem[\protect\citeauthoryear{Aharonian et~al.,}{Aharonian
  et~al.}{2006a}]{Aharonian:2006wh}
Aharonian F.,  et~al., 2006a, Phys. Rev. Lett., 97, 221102

\bibitem[\protect\citeauthoryear{Aharonian et~al.,}{Aharonian
  et~al.}{2006b}]{Aharonian:2006au}
Aharonian F.,  et~al., 2006b, Nature, 439, 695

\bibitem[\protect\citeauthoryear{{Aharonian} et~al.,}{{Aharonian}
  et~al.}{2008}]{2008A&A...492L..25A}
{Aharonian} F.,  et~al., 2008, Astron. Astrophys., 492, L25

\bibitem[\protect\citeauthoryear{Aharonian et~al.,}{Aharonian
  et~al.}{2009}]{GCSpectrum}
Aharonian F.,  et~al., 2009, Astron. Astrophys., 503, 817

\bibitem[\protect\citeauthoryear{{Aharonian} \& {Neronov}}{{Aharonian} \&
  {Neronov}}{2005a}]{Aharonian:2005ti}
{Aharonian} F.,  {Neronov} A.,  2005a, Astrophys. J., 619, 306

\bibitem[\protect\citeauthoryear{{Aharonian} \& {Neronov}}{{Aharonian} \&
  {Neronov}}{2005b}]{Aharonian:2005b}
{Aharonian} F.,  {Neronov} A.,  2005b, Astrophys. Space Science, 300, 255

\bibitem[\protect\citeauthoryear{Albert et~al.,}{Albert
  et~al.}{2006}]{Albert:2005kh}
Albert J.,  et~al., 2006, Astrophys. J., 638, L101

\bibitem[\protect\citeauthoryear{{Atoyan} \& {Dermer}}{{Atoyan} \&
  {Dermer}}{2004}]{Atoyan2004}
{Atoyan} A.,  {Dermer} C.~D.,  2004, Astrophys. J., 617, L123

\bibitem[\protect\citeauthoryear{{Bergstr{\"o}m}}{{Bergstr{\"o}m}}{2000}]{Berg%
stroem2000}
{Bergstr{\"o}m} L.,  2000, Rep. Progr. Phys., 63, 793

\bibitem[\protect\citeauthoryear{{B{\"u}sching} \& {de Jager}}{{B{\"u}sching}
  \& {de Jager}}{2008}]{Buesching2008}
{B{\"u}sching} I.,  {de Jager} O.~C.,  2008, Adv. Space Res., 42, 491

\bibitem[\protect\citeauthoryear{{B{\"u}sching}, {de Jager} \&
  {Snyman}}{{B{\"u}sching} et~al.}{2007}]{Buesching2007}
{B{\"u}sching} I.,  {de Jager} O.~C.,    {Snyman} J.,  2007, Astrophys. J.,
  656, 841

\bibitem[\protect\citeauthoryear{{Crocker} et~al.,}{{Crocker}
  et~al.}{2005}]{Crocker2005}
{Crocker} R.~M.,  et~al., 2005, Astrophys. J., 622, 892

\bibitem[\protect\citeauthoryear{{Dimitrakoudis}, {Mastichiadis} \&
  {Geranios}}{{Dimitrakoudis} et~al.}{2009}]{2009APh....31...13D}
{Dimitrakoudis} S.,  {Mastichiadis} A.,    {Geranios} A.,  2009, Astrop. Phys.,
  31, 13

\bibitem[\protect\citeauthoryear{{Fatuzzo} \& {Melia}}{{Fatuzzo} \&
  {Melia}}{2003}]{Fatuzzo2003}
{Fatuzzo} M.,  {Melia} F.,  2003, Astrophys. J., 596, 1035

\bibitem[\protect\citeauthoryear{{Green}}{{Green}}{2009}]{Green:2009qf}
{Green} D.~A.,  2009, Bull. Astron. Soc. India, 37, 45

\bibitem[\protect\citeauthoryear{Hinton \& Aharonian}{Hinton \&
  Aharonian}{2007}]{Hinton:2006zk}
Hinton J.~A.,  Aharonian F.~A.,  2007, Astrophys. J., 657, 302

\bibitem[\protect\citeauthoryear{{Hinton} \& {Hofmann}}{{Hinton} \&
  {Hofmann}}{2009}]{Hinton:2009}
{Hinton} J.~A.,  {Hofmann} W.,  2009, Ann. Rev. Astron. Astrophys., 47, 523

\bibitem[\protect\citeauthoryear{Kosack et~al.,}{Kosack
  et~al.}{2004}]{Kosack:2004ri}
Kosack K.,  et~al., 2004, Astrophys. J., 608, L97

\bibitem[\protect\citeauthoryear{{Kosack}}{{Kosack}}{2005}]{Kosack2005}
{Kosack} K.~P.,  2005, PhD thesis, Washington University, United States --
  Missouri

\bibitem[\protect\citeauthoryear{LaRosa et~al.,}{LaRosa
  et~al.}{2000}]{LaRosa00}
LaRosa T.,  et~al., 2000, Astron. J., 119, 207

\bibitem[\protect\citeauthoryear{{Liu} et~al.,}{{Liu}  et~al.}{2006}]{Liu2006a}
{Liu} S.,  et~al., 2006, Astrophys. J., 647, 1099

\bibitem[\protect\citeauthoryear{{Mizukami}}{{Mizukami}}{2008}]{Mizukami2008}
{Mizukami} T.,  2008, in {F.~A.~Aharonian, W.~Hofmann, \& F.~Rieger} ed., AIP
  Conf. Series Vol.~1085, {CANGAROO-III observation of gamma rays from the
  Galactic Center}.
p.~364

\bibitem[\protect\citeauthoryear{{Moore} et~al.,}{{Moore}
  et~al.}{1999}]{Moore1999}
{Moore} B.,  et~al., 1999, Mon. Not. R. Astron. Soc., 310, 1147

\bibitem[\protect\citeauthoryear{{Navarro}, {Frenk} \& {White}}{{Navarro}
  et~al.}{1997}]{Navarro1997}
{Navarro} J.~F.,  {Frenk} C.~S.,    {White} S.~D.~M.,  1997, Astrophys. J.,
  490, 493

\bibitem[\protect\citeauthoryear{Reid et~al.,}{Reid  et~al.}{1999}]{saga_radio}
Reid M.,  et~al., 1999, Astrophys. J., 524, 816

\bibitem[\protect\citeauthoryear{Tsuboi et~al.,}{Tsuboi  et~al.}{1999}]{cs}
Tsuboi M.,  et~al., 1999, Astrophys. J. Suppl., 120, 1

\bibitem[\protect\citeauthoryear{Tsuchiya et~al.,}{Tsuchiya
  et~al.}{2004}]{Tsuchiya:2004wv}
Tsuchiya K.,  et~al., 2004, Astrophys. J., 606, L115

\bibitem[\protect\citeauthoryear{Wang, Lu \& Gotthelf}{Wang
  et~al.}{2006}]{Wang:2005ya}
Wang Q.~D.,  Lu F.~J.,    Gotthelf E.~V.,  2006, Mon. Not. Roy. Astron. Soc.,
  367, 937

\bibitem[\protect\citeauthoryear{{Wang}, {Lu} \& {Chen}}{{Wang}
  et~al.}{2009}]{Wang:2009}
{Wang} Y.-P.,  {Lu} Y.,    {Chen} L.,  2009, Res. Astron. Astrophys., 9, 761

\bibitem[\protect\citeauthoryear{{Wommer}, {Melia} \& {Fatuzzo}}{{Wommer}
  et~al.}{2008}]{Wommer2008}
{Wommer} E.,  {Melia} F.,    {Fatuzzo} M.,  2008, Mon. Not. R. Astron. Soc.,
  387, 987

\bibitem[\protect\citeauthoryear{{Yusef-Zadeh} et~al.,}{{Yusef-Zadeh}
  et~al.}{1996}]{Yusef96}
{Yusef-Zadeh} F.,  et~al., 1996, Astrophys. J., 466, L25

\end{thebibliography}

\end{document}